\documentclass[10pt,conference]{IEEEtran}
\usepackage{psfrag}
\usepackage{subfigure}
\usepackage{cite}

\usepackage{array,multirow}
\usepackage{graphicx}  
\usepackage{psfrag}    
\usepackage{subfigure} 
\usepackage{amsmath}   
\usepackage{amssymb}   
\usepackage{eucal}     

\usepackage{pxfonts}
\usepackage{dsfont}

\DeclareMathAlphabet{\eurm}{U}{eur}{m}{n}
\DeclareMathAlphabet{\mathbsf}{OT1}{cmss}{bx}{n}
\DeclareMathAlphabet{\mathssf}{OT1}{cmss}{m}{sl}
\DeclareMathAlphabet{\mathcsf}{OT1}{cmss}{sbc}{n}

\DeclareSymbolFont{bsfletters}{OT1}{cmss}{bx}{n}  
\DeclareSymbolFont{ssfletters}{OT1}{cmss}{m}{n}
\DeclareMathSymbol{\bsfGamma}{0}{bsfletters}{'000}
\DeclareMathSymbol{\ssfGamma}{0}{ssfletters}{'000}
\DeclareMathSymbol{\bsfDelta}{0}{bsfletters}{'001}
\DeclareMathSymbol{\ssfDelta}{0}{ssfletters}{'001}
\DeclareMathSymbol{\bsfTheta}{0}{bsfletters}{'002}
\DeclareMathSymbol{\ssfTheta}{0}{ssfletters}{'002}
\DeclareMathSymbol{\bsfLambda}{0}{bsfletters}{'003}
\DeclareMathSymbol{\ssfLambda}{0}{ssfletters}{'003}
\DeclareMathSymbol{\bsfXi}{0}{bsfletters}{'004}
\DeclareMathSymbol{\ssfXi}{0}{ssfletters}{'004}
\DeclareMathSymbol{\bsfPi}{0}{bsfletters}{'005}
\DeclareMathSymbol{\ssfPi}{0}{ssfletters}{'005}
\DeclareMathSymbol{\bsfSigma}{0}{bsfletters}{'006}
\DeclareMathSymbol{\ssfSigma}{0}{ssfletters}{'006}
\DeclareMathSymbol{\bsfUpsilon}{0}{bsfletters}{'007}
\DeclareMathSymbol{\ssfUpsilon}{0}{ssfletters}{'007}
\DeclareMathSymbol{\bsfPhi}{0}{bsfletters}{'010}
\DeclareMathSymbol{\ssfPhi}{0}{ssfletters}{'010}
\DeclareMathSymbol{\bsfPsi}{0}{bsfletters}{'011}
\DeclareMathSymbol{\ssfPsi}{0}{ssfletters}{'011}
\DeclareMathSymbol{\bsfOmega}{0}{bsfletters}{'012}
\DeclareMathSymbol{\ssfOmega}{0}{ssfletters}{'012}

\newtheorem{theorem}{\textbf{Theorem}}
\newtheorem{corollary}{\textbf{Corollary}}

\newtheorem{definition}{\textbf{Definition}}

\newtheorem{remark}{Remark}
\newcommand{\dv}{\mathbf} 
\newcommand{\mc}{\mathcal} 

\begin{document}
\fontsize{9.4}{11.25pt}\selectfont

\title{Multiple Access Channel with States Known Noncausally at One Encoder and Only Strictly Causally at the Other Encoder\\}

\author{Abdellatif Zaidi$\:^{\dagger}$ \qquad Pablo Piantanida$\:^{\nmid}$ \qquad Shlomo Shamai (Shitz)$\:^{\ddagger}$\vspace{0.3cm}\\
$^{\dagger}$ Universit\'e Paris-Est Marne La Vall\'ee, Champs-sur-Marne 77454, France\\
$^{\nmid}$ Department of Telecommunications, SUPELEC, 91192 Gif-sur-Yvette, France\\
$^{\ddagger}$ Department of EE, Technion-Israel Institute of Technology, Haifa, Israel\\
abdellatif.zaidi@univ-mlv.fr, pablo.piantanida@supelec.fr, sshlomo@ee.technion.ac.il
}

\maketitle

\begin{abstract}
We consider a two-user state-dependent multiaccess channel in which the states of the channel are known non-causally to one of the encoders and only strictly causally to the other encoder. Both encoders transmit a common message and, in addition, the encoder that knows the states non-causally transmits an individual message. We study the capacity region of this communication model. In the discrete memoryless case, we establish inner and outer bounds on the capacity region. Although the encoder that sends both messages knows the states fully, we show that the strictly causal knowledge of these states at the other encoder can be beneficial for this encoder, and in general enlarges the capacity region. Furthermore, we find an explicit characterization of the capacity in the case in which the two encoders transmit only the common message. In the Gaussian case, we characterize the capacity region for the model with individual message as well. Our converse proof in this case shows that, for this model, strictly causal knowledge of the state at one of the encoders does not increase capacity if the other is informed non-causally, a result which sheds more light on the utility of conveying a compressed version of the state to the decoder in recent results by Lapidoth and Steinberg on a multiacess model with only strictly causal state at both encoders and independent messages.
\end{abstract}

\section{Introduction}\label{secI}


Multiple access channels with states known causally at the encoders  have been studied recently in \cite{LS10a, LS10b} and \cite{LSY10} (see also \cite{BL10,J06,SK05}). In \cite{LS10a}, the states are known in a strictly causal manner at both encoders which transmit independent messages. The authors show that the strict knowledge of the states can be beneficial, in the sense that it increases the capacity for this model. This result is reminiscent of Dueck's proof \cite{D80} that feedback can increase the capacity region of some broadcast channels. In accordance with \cite{D80}, the main idea of the achievability result in \cite{LS10a} is a block Markov coding scheme in which the two users collaborate to describe the state to the decoder by sending cooperatively a compressed version of it. As noticed in \cite{LS10a}, although some non-zero rate that otherwise could be used to transmit pure information is spent in describing the state to the decoder, the net effect can be an increase in the capacity.

In \cite{LS10a} and \cite{LS10b}, an encoder that benefits from the availability of states at the other encoder (strictly causally) does not know the states fully itself, i.e., it knows the states only strictly causally itself. One can then wonder whether, in a multiaccess channel, the knowledge of the states only strictly causally at one encoder could be of any help to another encoder which knows the states non-causally.

In this paper, we study a two-user state-dependent multiple access channel with the channel states known non-causally at one encoder and only strictly causally at the other encoder. Both encoders transmit a common message and, in addition, the encoder that knows the states non-causally transmits an individual message. This model generalizes one whose capacity region is established in \cite{SBSV07a} and in which the encoder that sends only the common message does not know the states at all. More precisely, let $W_c$ and $W_1$ denote the common message and the individual message to be transmitted in, say, $n$ uses of the channel; and $S^n=(S_1,\hdots,S_n)$ denote the state sequence affecting the channel during this time. At time $i$, Encoder 1 knows the complete sequence $S^n=(S_1,\hdots,S_{i-1},S_i,\hdots,S_n)$ and sends $X_{1i}=\phi_1(W_c,W_1,S^n)$, and Encoder 2 knows \textit{only} $S^{i-1}=(S_1,\hdots,S_{i-1})$ and sends $X_{2i}=\phi_{2,i}(W_c,S^{i-1})$ -- the functions $\phi_1$ and $\phi_{2,i}$ are some encoding functions.

We study the capacity region of this state-dependent MAC model. In the discrete memoryless case, we establish an inner bound and an outer bound on the capacity region of this model. The achievable region is based on a coding scheme that generalizes the Gelf'and-Pinsker binning-like scheme of \cite{SBSV07a} by letting the two encoders also collaborate to send a description of the state to the decoder through Wyner-Ziv compression \cite{WZ76}, in the spirit of \cite{LS10a} and \cite{LS10b}. By studying a special case, we show that this can be beneficial in general, even for the encoder that knows the states non-causally. More specifically, we show that, even though it knows the states fully, this encoder can still get benefit from the strictly causal knowledge of these states at the other encoder. Equivalently, this shows that the capacity region of the DM model that we study is strictly bigger than that of the model in \cite{SBSV07a}.

For the case in which both encoders send only the common message, we characterize the capacity of our model. Also, we show that the knowledge of the states only strictly causally does \textit{not} increase the capacity. We note that the non-utility of the strictly causal availability of the states in this case is not a direct consequence of that of the non-utility of these states for the model \cite{LS10a} in the case in which both transmitters send only a common message.


Furthermore, we also study a memoryless Gaussian model in which both the noise and the state are additive and Gaussian. In this case, we characterize the capacity region of our model and show that, in contrast to \cite{LS10a}, the availability of the states in a strictly causal manner at the encoder that sends only the common message is of no utility; or, equivalently, one does no better than had this encoder known only the common message.

\section{Problem Setup}\label{secII}

\vspace{0.2cm}

We consider a stationary memoryless state-dependent MAC $W_{Y|X_1,X_2,S}$  whose output $Y \in \mc Y$ is controlled by the channel inputs $X_1 \in \mc X_1$ and $X_2 \in \mc X_2$ from the encoders and the channel state $S \in \mc S$ which is drawn according to a memoryless probability law $Q_S$. We assume that the channel state $S^n$ is known non-causally at Encoder 1, i.e., beforehand, at the beginning of the transmission block. Encoder 2 knows the channel states only strictly-causally; that is, at time $i$, it knows the states only up to time $i-1$, $S^{i-1}=(S_1,\hdots,S_{i-1})$.

Encoder 2 wants to send a common message $W_c$ and Encoder 1 wants to send an independent individual message $W_1$ along with the common message $W_c$. We assume that the common message $W_c$ and the individual message $W_1$ are independent random variables drawn uniformly from the sets $\mc W_c=\{1,\cdots,M_c\}$ and  $\mc W_1=\{1,\cdots,M_1\}$, respectively. The sequences $X_{1}^n$ and $X_{2}^n$ from the encoders are sent across a state-dependent multiple access channel modeled as a memoryless conditional probability distribution $W_{Y|X_1,X_2,S}$. The joint probability mass function on ${\mc W_c}{\times}{\mc W_1}{\times}{\mc S^n}{\times}{\mc X^n_1}{\times}{\mc X^n_2}{\times}{\mc Y^n}$ is given by
\begin{align}
& P(w_c,w_1,s^n,x^n_1,x^n_2,y^n) = P(w_c)p(w_1)\prod_{i=1}^{n}\Big[Q_S(s_i)P(x_{1,i}|w_c,w_1,s^n)\nonumber\\
&\hspace{2.5cm} {\cdot}P(x_{2,i}|w_c,s^{i-1}){\cdot}W_{Y|X_1,X_2,S}(y_i|x_{1,i},x_{2,i},s_i)\Big].
\end{align}
The receiver guesses the pair $(\hat{W}_c,\hat{W}_1)$ from the channel output $Y^n$.

\begin{definition}
For positive integers $n$, $M_c$ and $M_1$, an $(M_c,M_1,n,\epsilon)$ code for the multiple access channel with states known noncausally at one encoder and only strictly causally at the other encoder consists of a mapping
\begin{align*}
\phi_1: \mc W_c{\times}\mc W_1{\times}\mc S^n \longrightarrow \mc X^n_1
\end{align*}
at Encoder 1, a sequence of mappings
\begin{align*}
\phi_{2,i}: \mc W_c{\times}\mc S^{i-1} \longrightarrow \mc X_2, \quad i=1,\hdots,n
\end{align*}
at Encoder 2, and a decoder map
\begin{align*}
\psi : \mc Y^n \longrightarrow \mc W_c{\times}\mc W_1
\end{align*}
such that the average probability of error is bounded by $\epsilon$,
\begin{equation*}
P_e^n = \mathbb{E}_{S}\big[\mathrm{Pr}\big(\psi(Y^n)\neq (W_c,W_1)|S^n=s^n\big)\big] \leq \epsilon.
\end{equation*}
The rate of the common message and the rate of the individual message are defined as
\begin{align*}
&R_c = \frac{1}{n}\log M_c \qquad \text{and} \qquad R_1 = \frac{1}{n}\log M_1,
\end{align*}
respectively.
\end{definition}

A rate pair $(R_c,R_1)$ is said to be achievable if for every $\epsilon > 0$ there exists an $(2^{nR_c},2^{nR_1},n,\epsilon)$ code for the channel $W_{Y|X_1,X_2,S}$.  The capacity region of the considered state-dependent MAC is defined as the closure of the set of achievable rate pairs.

Due to space limitation, the results of this paper are either outlined only or mentioned without proofs. Detailed proofs and improved results for the model of this paper can be found in \cite{ZPS11b}.

\vspace{-0.3cm}

\section{The Discrete Memoryless Model}

In this section, it is assumed that the alphabets $\mc S, \mc X_1, \mc X_2$ are finite.

\subsection{Bounds on the Capacity Region}\label{secIII_subsecA}

The following Theorem provides an inner bound on the capacity region of the state-dependent discrete memoryless MAC model that we study. An ouline of its proof is given in Section~\ref{secV}. 

\begin{theorem}\label{Theorem__InnerBoundDiscreteMemorylessChannel}
The capacity region of the multiple access channel with states known non-causally at one encoder and strictly causally at the other encoder contains the closure of the set of all rate-pairs $(R_c,R_1)$ satisfying
\begin{align}
R_1 \: &\leq \: I(U;Y|V,X_2)-I(U;S|V,X_2) \nonumber\\
R_c+ R_1 \: &\leq \: I(U,V,X_2;Y)-I(U,V,X_2;S),
\label{InnerBoundDiscreteMemorylessChannel}
\end{align}
for some probability distribution of the form
\begin{align}
&P_{S,V,U,X_1,X_2,Y}=Q_SP_{X_2}P_{V|S}P_{U,X_1|S,X_2}W_{Y|X_1,X_2,S}
\label{MeasureForInnerBoundDiscreteMemorylessChannel}
\end{align}
and satisfying
\begin{align}
I(V,X_2;Y)-I(V,X_2;S) &\geq 0.
\label{NonNegativityConstraint}
\end{align}
\end{theorem}

\vspace{0.2cm}

\begin{remark}
\noindent The joint distribution \eqref{MeasureForInnerBoundDiscreteMemorylessChannel} satisfies the Markov relation $V \leftrightarrow S \leftrightarrow (U,X_1,X_2,Y)$, and that $X_2$ is independent of $(S,V)$. Also, it is insightful to note that the region \eqref{InnerBoundDiscreteMemorylessChannel} can be written as
\begin{align}
R_1 \: &\leq \: I(U;Y|V,X_2)-I(U;S|V,X_2) \nonumber\\
(R_c+R_s)+ R_1 \: &\leq \: I(U,X_2;Y|V)-I(U,X_2;S|V),
\label{EquivalentFormInnerBoundDiscreteMemorylessChannel}
\end{align}
where $R_s = I(V;S)-I(V;Y) \geq 0$.
\end{remark}

\vspace{0.2cm}

\begin{remark}
\noindent The inner bound is based on a coding scheme in which a lossy version of the state is conveyed to the decoder using Wyner-Ziv compression \cite{WZ76} and block-Markov encoding, in the spirit of \cite{LS10a} and \cite{LS10b}, combined with a generalized Gelfand-Pinsker binning scheme for the MAC with states known non-causally at only the encoder which sends both messages and no states at the encoder which sends only the common message \cite{SBSV07a}. More specifically, fix a measure $P_{S,U,V,X_1,X_2,Y}$ satisfying \eqref{MeasureForInnerBoundDiscreteMemorylessChannel}. The transmission is performed in $B+1$ blocks. We denote by $\dv s[i]$ the channel state in block $i$, $i=1,\hdots,B+1$. In each block $i$, the encoder that knows the states only strictly causally, i.e., Encoder 2, sends  to the decoder a compressed version $\dv v_i$ of the state $\dv s[i-1]$ that affects the previous block $i-1$ through Wyner-Ziv compression at rate $R_s=I(V;S)-I(V;Y)$. The goal is that, when decoded at the decoder before the information messages of block $i-1$, this information can be used as side information at the decoder, thus improving the corresponding rates. Then, accounting for the fact that the decoder will actually know $\dv v_{i+1}$ for the decoding in block $i$,  in block $i$ Encoder 2 can transmit at rate $R_2=I(X_2;Y|V)$. This rate can be shared among sending to the decoder the Wyner-Ziv compressed version $\dv v_i$ of the state $\dv s[i-1]$ at rate $R_s=I(V;S)-I(V;Y)$, and pure information for the common message  at the remaining rate $R'_2=I(V,X_2;Y)-I(V,X_2;S)$. Now, the encoder that knows the states non-causally, i.e., Encoder 1, can transmit using a Gel'fand-Pinsker-like scheme, at rate $R'_1=I(U;Y,V,X_2)-I(U;S,V,X_2)=I(U;Y|V,X_2)-I(U;S|V,X_2)$, by treating $(V,X_2)$ as part of the state information at the encoder and accounting for the fact that this state will also be available at the decoder once the information from Encoder 2 has been decoded first. Finally, the information sent by Encoder 1 at rate $R'_1$ can be shared among the private message $W_1$ and the common message $W_c$.
\end{remark}

\vspace{0.2cm}

We now establish an outer bound on the capacity region of the DM MAC model that we study. In the next section, we will show that this outer bound is actually tight for the Gaussian model.

\begin{theorem}\label{Theorem__AlternativeOuterBoundDiscreteMemorylessChannel}

The capacity region of the multiple access channel with states known non-causally at one encoder and strictly causally at the other encoder is contained in the closure of the set of all rate-pairs $(R_c,R_1)$ satisfying
\begin{align}
R_1 \: &\leq \: I(X_1;Y|S,X_2), \nonumber\\
R_c+ R_1 \: &\leq \: I(X_1,X_2;Y|S)-I(X_2;S|Y),
\label{AlternativeOuterBoundDiscreteMemorylessChannel}
\end{align}
for some probability distribution of the form
\begin{align}
&P_{S,X_1,X_2,Y}=Q_SP_{X_2}P_{X_1|X_2,S}W_{Y|X_1,X_2,S}.
\label{MeasureForAlternativeOuterBoundDiscreteMemorylessChannel}
\end{align}
\end{theorem}


\vspace{0.2cm}

\subsection{Common-message Capacity}\label{secIII_subsecC}

In this section, we characterize the capacity in the case in which the two encoders transmit only the common message, i.e., $R_1=0$. We refer to it as \textit{common-message capacity}.

\vspace{0.2cm}

\begin{theorem}\label{Theorem__CommonMessageCapacity}
The common message capacity, $\mc C$, of the multiple access channel with common message and states known non-causally at one encoder and strictly causally at the other encoder is given by
\begin{align}
\mc C &= \max I(U,X_2;Y)-I(U;S|X_2)
\end{align}
where the maximization is over joint measures $P_{S,U,X_1,X_2,Y}$  of the form
\begin{align}
P_{S,U,X_1,X_2,Y} &= Q_SP_{X_2}P_{U,X_1|S,X_2}.
\end{align}
\end{theorem}


\vspace{0.2cm}

\begin{remark}\label{remark5}
The capacity of our model in Theorem~\ref{Theorem__CommonMessageCapacity} is the same as the one of the model with state $S^n$ at Encoder 1 and no state at all at Encoder 2 established in \cite{SBSV07a}. This shows that the strictly causal knowledge of the state at Encoder 2 does not increase capacity. The non-utility of the strictly causal availability of the states in this model is not a direct consequence of that of the non-utility of these states for the model \cite{LS10a} in which both transmitters send only a common message. Also, in contrast to \cite{SBSV07a} and also \cite{ZKLV09a}, our converse proof does not follow directly from the converse part proof of the capacity formula for the standard Gelf'and-Pinsker channel \cite{GP80} because, at time $i$, Encoder 2 sends inputs which are function of not only the message to transmit, but also the past state sequence $S^{i-1}$. The converse proof includes a redefinition of the involved auxiliary random variable.
\end{remark}

\vspace{-0.3cm}

\subsection{Example: Private Message Only}\label{secIII_subsecB}
In this section, we use $h(\alpha)$ to denote the entropy of a Bernoulli\:$(\alpha)$ source, i.e.,
\begin{equation}
h(\alpha) = - \alpha \log(\alpha) - (1-\alpha)\log(1-\alpha)
\end{equation}
and $p * q$ to denote the binary convolution, i.e.,
\begin{equation}
p * q = p(1-q)+q(1-p).
\end{equation}

\begin{figure}[htpb]
\centering
\includegraphics[width=0.8\linewidth]{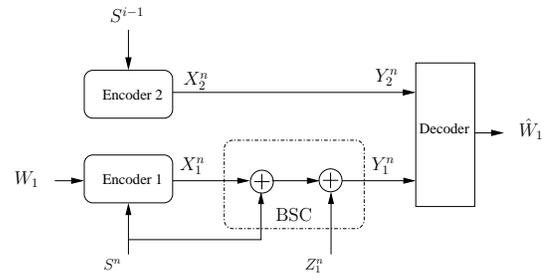}
\caption{Binary state-dependent MAC example with two output components, $Y^n=(Y^n_1,Y^n_2)$, with $Y^n_1=X^n_1 + S^n + Z^n_1$ and $Y^n_2=X^n_2$.}
\label{BSCModelCounterExample}
\end{figure}

Consider the binary memoryless MAC shown in Figure~\ref{BSCModelCounterExample}. Here, all the random variables are binary $\{0,1\}$. The channel has two output components, i.e., $Y^n=(Y^n_1,Y^n_2)$. The component $Y^n_2$ is deterministic, $Y^n_2=X^n_2$, and the component $Y^n_1=X^n_1 + S^n + Z^n_1$, where the addition is modulo $2$. Encoder 2 knows the states only strictly causally and has no message to transmit. Encoder 1 knows the states non-causally and transmits an individual message $W_1$. The state and noise vectors are independent and memoryless, with the state process $S_i$, $i \geq 1$, and the noise process $Z_{1,i}$, $i \geq 1$, assumed to be Bernoulli $(\frac{1}{2})$ and  Bernoulli $(p)$ processes, respectively. The vectors $X^n_1$ and $X^n_2$ are the channel inputs, subjected to the constraints
\vspace{-0.3cm}
\begin{align}
\sum_{i=1}^{n} X_{1,i} &\leq nq_1 \quad \text{and} \quad \sum_{i=1}^{n} X_{2,i} \leq nq_2.
\label{BinaryChannel__InputsConstraints}
\end{align}
\vspace{-0.2cm}

For this example, as we will show shortly, the strictly causal knowledge of the states at Encoder 2 \textbf{does} help, and in fact Encoder 1 can transmit at rates that are larger than the standard Gelf'and-Pinsker rate $I(U;Y_1)-I(U;S)$ which would be the capacity had Encoder 2 been of no help.

\vspace{0.2cm}

\textit{Claim 1:} The capacity of the state-dependent binary memoryless MAC shown in Figure~\ref{BSCModelCounterExample} is given by
\begin{align}
\mc C &= \max_{p(x_1|s)} \:\: I(X_1;Y_1|S).
\label{Capacity__ModelCounterExample}
\end{align}

\vspace{0.1cm}

\textit{Proof:} 1) The achievability follows from Theorem~\ref{Theorem__InnerBoundDiscreteMemorylessChannel}, by setting $R_c=0$, $V=S$, $U=X_1$, $Y_2=X_2$ with $X_2 \sim$ Bernoulli\:$(\frac{1}{2})$ independent of $(S,X_1)$.

2) The converse follows straightforwardly by specializing the cut-set upper bound to this example
\begin{align}
R &\leq I(X_1;Y|X_2,S)\\
  &= I(X_1;Y_1|X_2,S)\\
  &= H(Y_1|X_2,S)-H(Y_1|X_1,X_2,S)\\
  \label{ProofCutSetBoundCounterExample__Step1}
  &\leq H(Y_1|S)-H(Y_1|X_1,X_2,S)\\
  &\leq  H(Y_1|S)-H(Y_1|X_1,S)\\
  \label{ProofCutSetBoundCounterExample__Step2}
  &= I(X_1;Y_1|S),
\end{align}
where \eqref{ProofCutSetBoundCounterExample__Step1} holds since condutioning reduces entropy, and \eqref{ProofCutSetBoundCounterExample__Step2} holds by the Markov relation $X_2 \leftrightarrow (X_1,S) \leftrightarrow Y_1$.

\vspace{0.2cm}

\textit{Claim 2:} The capacity of the state-dependent binary memoryless MAC shown in Figure~\ref{BSCModelCounterExample} satisfies
\begin{align}
\mc C &= h(p * q_1) - h(p) > \max_{p(u,x_1|s)} \:\: I(U;Y_1)-I(U;S).
\label{ExplicitCharacterization__Capacity__ModelCounterExample}
\end{align}

\vspace{0.2cm}

\textit{Proof:} Claim 2 is a simple consequence of Claim 1 and known results on the capacity of the binary dirty paper channel (see for example \cite{PCR03} and references therein).

\vspace{0.2cm}

\begin{remark}
In this example, the encoder that knows the states only strictly causally simply conveys these states to the receiver, noiselessly. The receiver then becomes aware of the channel states fully (since the delay in learning these states at the decoder has no impact on the capacity). This explains why Encoder 1 can transmit at rates that can be strictly larger than the standard Gelfand-Pinker rate $\max_{p(u,x_1|s)} \:\: I(U;Y_1)-I(U;S)$; and in fact achieves the capacity \eqref{ExplicitCharacterization__Capacity__ModelCounterExample} of a state-dependent additive binary channel with the states known at both transmitter and receiver ends.
\end{remark}

\section{Gaussian Model}

In this section, we consider a two-user state-dependent Gaussian MAC in which the channel states $S^n$ and the noise are additive and Gaussian. As in Section \ref{secII}, we assume that Encoder 1 knows the channel states non-causally and Encoder 2 knows the channel states strictly causally. The two encoders send some common message $W_c$; and, in addition, Encoder 1 sends an individual message $W_1$.

At time instant $i$, the channel output $Y_i$ is related to channel inputs $X_{1,i}$ and $X_{2,i}$ from the two encoders, the channel state $S_i$ and the noise $Z_i$ by
\begin{align}
& Y_i=X_{1,i}+X_{2,i}+S_i+Z_i,
\label{ChannelModelForGaussianMACWithAsymmetricCSI}
\end{align}
where $S_i$ and $Z_i$ are zero-mean Gaussian random variables with variance $Q$ and $N$, respectively. The random variables $S_i$ and $Z_i$ at time instant $i \in \{1,\cdots,n\}$ are mutually independent, and independent from $(S_j,Z_j)$ for $j \neq i$. Also, at time $i$, the input $X_{2,i}$ is independent from the state $S_i$.

We consider the individual power constraints on the transmitted power
\begin{equation}
\sum_{i=1}^{n}X_{1,i}^2 \leq nP_1, \:\: \sum_{i=1}^{n}X_{2,i}^2 \leq nP_2.
\label{IndividualPowerConstraintsFullDuplexRegime}
\end{equation}
The definition of a code for this channel is the same as given in Section \ref{secII}, with the additional power constraint \eqref{IndividualPowerConstraintsFullDuplexRegime}.

\noindent The following theorem provides the capacity region of the studied Gaussian model.

\vspace{0.2cm}

\begin{theorem}\label{Theorem__CapacityRegionMemorylessGaussianChannel}
 The capacity region of the Gaussian model \eqref{ChannelModelForGaussianMACWithAsymmetricCSI} is given by the set of all the rate pairs $(R_c,R_1)$ satisfying
 \begin{align}
 R_1\: &\leq \:\frac{1}{2}\log\Big(1+\frac{P_1(1-\rho^2_{12}-\rho^2_{2s})}{N}\Big)\nonumber\\
 R_c+R_1 \: & \leq \frac{1}{2}\log\Big(1+\frac{(\sqrt{P_2}+\rho_{12}\sqrt{P_1})^2}{P_1(1-\rho^2_{12}-\rho^2_{1s})+(\sqrt{Q}+\rho_{1s}\sqrt{P_1})^2+N}\Big)\nonumber\\&+\frac{1}{2}\log\Big(1+\frac{P_1(1-\rho^2_{12}-\rho^2_{1s})}{N}\Big),
 \label{OuterBoundGaussianChannel}
 \end{align}
 where the maximization is over $\rho_{12} \in [0,1]$, $\rho_{1s} \in [-1,0]$ such that
 \begin{equation}
 \rho^2_{12}+\rho^2_{1s} \leq 1.
 \label{AllowableCovarianceMatrixOuterBound}
 \end{equation}
 \end{theorem}


 \vspace{0.2cm}

 \begin{remark}\label{remark3}
 The capacity region of our model in Theorem~\ref{Theorem__CapacityRegionMemorylessGaussianChannel} is the same as the one of the model with state $S^n$ at Encoder 1 and no state at all at Encoder 2 established in \cite[Theorem 7]{SBSV07a}. Our converse proof then proves that, for our model, it is \textit{optimal} to just ignore the known $S^{i-1}$ at Encoder 2 and use the coding scheme of \cite{SBSV07a}. That is, one can do no better exploitation of the state $S^{i-1}$ at Encoder 2. While one could expect some utility of the collaborative transmission of  $S^{i-1}$ as in the Gaussian setup in Lapidoth and Steinberg \cite{LS10a}, a direct consequence of our converse proof is that this would be of no help (in the sense that it would not result in a better transmission rate).

\noindent This can be explained as follows. As it can be seen from the proof of Theorem~\ref{Theorem__InnerBoundDiscreteMemorylessChannel}, the joint transmission of the state $\dv s[i-1]$ in block $i$ aims at equipping the decoder with an estimate of the state, which is then utilized as decoder side information for decoding the messages in block $i-1$. In general, this can be beneficial as we already mentioned. In the Gaussian case, however, Encoder 1 knows the state non-causally here and can cancel its effect completely (for the transmission of the private message) using a variation of the standard dirty paper scheme \cite{C83}, with no need to diminishing its effect via the joint transmission of the compressed version of the state. 


 \end{remark}

\vspace{0.2cm}

\noindent The following corollary follows straightforwardly from Theorem~\ref{Theorem__CapacityRegionMemorylessGaussianChannel}.

\begin{corollary}\label{corollary1}
The common message capacity, $\mc C_{\text{G}}$, of the Gaussian model \eqref{ChannelModelForGaussianMACWithAsymmetricCSI} is given by
  \begin{align}
  \mc C\: = \max\: & \frac{1}{2}\log\Big(1+\frac{(\sqrt{P_2}+\rho_{12}\sqrt{P_1})^2}{P_1(1-\rho^2_{12}-\rho^2_{1s})+(\sqrt{Q}+\rho_{1s}\sqrt{P_1})^2+N}\Big)\nonumber\\
& +\frac{1}{2}\log\Big(1+\frac{P_1(1-\rho^2_{12}-\rho^2_{1s})}{N}\Big),
  \label{OuterBoundGaussianChannel}
  \end{align}
  where the maximization is over $\rho_{12} \in [0,1]$, $\rho_{1s} \in [-1,0]$ such that
  \begin{equation}
  \rho^2_{12}+\rho^2_{1s} \leq 1.
  \label{AllowableCovarianceMatrixCommonMessageCapacity}
  \end{equation}
  \end{corollary}

\section{Outline of Proof of Theorem~\ref{Theorem__InnerBoundDiscreteMemorylessChannel}}\label{secV}

In the following, we show that the rate pair
\begin{align} 
R_c &= I(V,X_2;Y)-I(V,X_2;S)\nonumber\\
R_1 &= I(U;Y|V,X_2)-I(U;S|V,X_2)
\label{CornerPoint__InnerBoundDiscreteMemorylessChannel}
\end{align}
is achievable. The complete region in Theorem~\ref{Theorem__InnerBoundDiscreteMemorylessChannel} will then be achievable by sharing the rate $R_1$ in \eqref{CornerPoint__InnerBoundDiscreteMemorylessChannel} among sending the individual information and additional common information as we indicated in the aforementioned remark.

\noindent First we generate a random codebook that we use to obtain the rate-pair~\eqref{CornerPoint__InnerBoundDiscreteMemorylessChannel}. Next, we outline the encoding and decoding procedures.

Note that the rate $R_1$ in \eqref{CornerPoint__InnerBoundDiscreteMemorylessChannel} can be written as $R_1=I(U;Y,V|X_2)-I(U;S|X_2)$.  We transmit in $B+1$ blocks, each of length $n$. During each of the first $B$ blocks, both encoders transmit a message $w_{c,i} \in [1,2^{nR_c}]$, and Encoder 1 also sends an individual message $w_{1,i} \in [1,2^{nR_1}]$, where $i=1,\hdots,B$ denotes the index of the block. For fixed $n$, the average rate-pair $(R_c\frac{B}{B+1},R_1\frac{B}{B+1})$ over $B+1$ blocks approaches $(R_c,R_1)$ as $B \longrightarrow +\infty$.

\noindent \textbf{Codebook Generation:} Fix a measure $P_{S,U,V,X_1,X_2,Y}$ of the form \eqref{MeasureForInnerBoundDiscreteMemorylessChannel} and satisfying \eqref{NonNegativityConstraint}. Fix $\epsilon > 0$ and denote $M_c = 2^{n[I(V,X_2;Y)-I(V,X_2;S)-\epsilon]}$,
\begin{align}
M_V &= 2^{n[I(V;S)-I(V;Y)-\epsilon]}\qquad & J_V &= 2^{n[I(V;Y)+2\epsilon]}\nonumber\\
M_1&= 2^{n[I(U;Y,V|X_2)-I(U;S|X_2)-4\epsilon]}\qquad & J_U &= 2^{n[I(U;S|X_2)+2\epsilon]}.
\label{ValuesForBinningVariablesInTheorem1}
\end{align}

\begin{itemize}
\item[1)] We generate $J_VM_V$ independent and identically distributed (i.i.d.) codewords $\dv v(m,j_V)$ indexed by $m=1,\hdots,M_V$, $j_V=1,\hdots,J_V$, each with i.i.d. components drawn according to $P_{V}$.
\item[2)] Independently, we generate $M_VM_c$ i.i.d. codewords $\dv x_2(m,l)$ indexed by  $m=1,\hdots,M_V$, $l=1,\hdots,M_c$, each with i.i.d. components drawn according to $P_{X_2}$.
\item[3)] For each codeword $\dv x_2(m,l)$, we generate a collection of $J_UM_1$ i.i.d. codewords $\{\dv u(m,l,k,j)\}$ indexed by $k=1,\hdots,M_1$, $j_U=1,\hdots,J_U$, each with i.i.d. components draw according to $P_{U|X_2}$.
\end{itemize}

\textbf{Encoding:}
Suppose that a common message $W_c=l$ and an individual message $W_1=k$ are to be transmitted. The message $W_c$ is divided into $B$ blocks $(w_{c,1},w_{c,2},\hdots,w_{c,B})$ of $nR_c$ bits each, and the  message $W_1$ is divided into $B$ blocks $(w_{1,1},w_{1,2},\hdots,w_{1,B})$ of $nR_{1}$ bits each. For convenience we let $w_{c,B+1}=w_{1,B+1}=1$. The transmission is performed in $B+1$ blocks. We denote by $\dv s[i]$ the channel state in block $i$, $i=1,\hdots,B+1$. Let $m_1=1$ (a default value) and the bin index $m_{B+1}$ be selected such that $\dv v(m_{B+1},j_{V_{B+1}})$ is strongly jointly with $\dv s[B]$ for some $j_{V_{B+1}} \in \{1,\hdots,J_V\}$ (with probability near one, there will exist such bin index). We assume that $m_{B+1}$ is known at the decoder and so will not be transmitted in the last block (this does not alter the rate for large $B$).

Continuing with the strategy, let $(l_i,k_i)$ be the new common and individual messages to be sent at the beginning of block $i$. Encoder $2$ looks for an index $m_i$  such that $\dv v(m_i,j_{V_i})$ is strongly jointly with $\dv s[i-1]$ for some $j_{V_i} \in \{1,\hdots,J_V\}$. If there is no such vector $\dv v$, $m_i$ is set to $1$ and an error is declared. Encoder 2 then transmits the vector $\dv x_2(m_i,l_i)$. To transmit the pair $(l_i,k_i)$, Encoder 1 first looks for the smallest $j_{Ui}$ such that $\dv u(m_i,l_i,k_i,j_{Ui})$ is jointly typical with $(\dv x_2(m_i,l_i),\dv s[i])$. Denote this $j_{Ui}$ by $j^{\star}_{Ui}=j_U(\dv s[i],m_i,l_i,k_i)$. If such $j^{\star}_{Ui}$ is not found, or if the observed state is not typical, an error is declared and $j_U(\dv s[i],m_i,l_i,k_i)$ is set to $j_{Ui}=J_U$. Notice that, if it exists, the chosen $\dv u(m_i,l_i,k_i,j^{\star}_{Ui})$ will be jointly typical with $(\dv x_2(m_i,l_i),\dv s[i],\dv v(m_{i+1},j_{V_{i+1}}))$, where $\dv v(m_{i+1},j_{V_{i+1}})$ is the covering codeword of the next block, selected such that $\dv v(m_{i+1},j_{V_{i+1}})$ is strongly jointly with $\dv s[i]$. Encoder 1 then transmits a vector $\dv x_1[i]$ which is drawn i.i.d. conditionally given $\Big(\dv s[i], \dv x_2(m_i,l_i), \dv u(m_i,l_i,k_i,j^{\star}_{Ui})\Big)$ (using the conditional measure $P_{X_1|S,U,X_2}$ induced by  \eqref{MeasureForInnerBoundDiscreteMemorylessChannel}).

\textbf{Decoding:} The decoding procedure at the destination is based on a combination of joint typicality and backward-decoding.

\noindent At the end of the transmission, the destination has collected all the blocks of channel outputs $\dv y_3[1],\hdots,\dv y_3[B+1]$, and can then perform Willems' backward-decoding by first decoding $(m_B,l_B,k_B)$ from $\dv y_3[B]$.

\noindent Decoding the information of block $B$ is as follows. First, the decoder knows the bin index $m_{B+1}$ and looks for a codeword index $\hat{j}_{VB+1}$ such that the covering codeword $\dv v(m_{B+1},\hat{j}_{VB+1})$ inside this bin is strongly jointly typical with the channel output $\dv y[B]$ of block $B$. If there is no such $\hat{j}_{VB+1}$, or more than one such $\hat{j}_{VB+1}$, the decoder sets $\hat{j}_{VB+1}$ to $J_V$ and declares an error. Next, the decoder declares that $(\hat{l}_B,\hat{k}_B)$ is sent if there is a unique triple $(\hat{m}_B,\hat{l}_B,\hat{k}_B)$ such that $\dv x_2(\hat{m}_B,\hat{l}_B)$, $\dv u(\hat{m}_B,\hat{l}_B,\hat{k}_B,j_{UB})$ are jointly typical with the augmented output $(\dv y_3[B],\dv v(m_{B+1},j_{VB+1}))$, for some $j_{UB} \in \{1,\hdots,J_U\}$. One can show that, with the choice \eqref{ValuesForBinningVariablesInTheorem1}, the decoding error in this step is small for sufficiently large $n$.


\section*{Acknowledgement}
\vspace{0.1cm}

Insightful discussions with H. Permuter are gratefully acknowledged. This work has been supported by the European Commission in the framework of the FP7 Network of Excellence in Wireless Communications (NEWCOM++). The work of S. Shamai has also been supported by the CORNET consortium.

\bibliographystyle{IEEEtran}
\bibliography{paperISIT2011}
\end{document}